\documentclass{ws-p8-50x6-00}

\begin{document}
\title{The Helium abundance problem and non-minimally coupled quintessence}

\author{Xuelei Chen}

\address{Physics Department, Ohio State University, \\
174 W18th Ave., Columbus, OH ~43210, 
USA\\
E-mail: xuelei@pacific.mps.ohio-state.edu}

 
\maketitle

\abstracts{There is a tension between observed Helium abundance 
and the prediction of the standard Big Bang Nucleosynthesis. We show that
non-minimally quintessence model may help to reduce this tension
between theory and observation.}

Recently, it has been discovered that the expansion of
the Universe is accelerating\cite{SNIa}. This requires the existence
of a dark energy component in the Universe with an equation of state
$p = w \rho$, $w<0$. One example of such a component is a
cosmological constant. However, it is difficult to understand in the 
present framework of particle physics why it is 
so small. A fine tuning of $10^{-120}$ is required
if the cosmological constant arises from Planck scale physics.

Quintessence models\cite{quintessence} were suggested as an
alternative to cosmological constant. 
In quintessence models, a scalar field provides the dark energy
which drives the accelerated expansion of the Universe.
The evolution of the scalar field is such that its equation of state
mimics the dominant component of the Universe,
thus explains why it is so small at present time.

In Non-minimally Coupled (NMC) quintessence models\cite{NMC} 
the scalar field couples to the gravitational constant. The action of
NMC can be written as 
\begin{equation}
S=\int d^{4} x \sqrt{-g} \left[\frac{1}{2}F R-\frac{1}{2}\phi^{;\mu} 
\phi_{;\mu} -V(\phi) + L_{\rm fluid}\right],\nonumber 
\label{NMCeq}
\end{equation}
where
\begin{equation}
F(\phi)=1-\xi(\phi^2 - \phi_0^{2}),\quad
V(\phi)=V_{0} \phi^{-\alpha}.
\end{equation}
Such coupling could arise if for example, 
the scalar field is a dilaton in superstring theory. 
One of the motivations for introducing 
such non-minimal coupling is to address the ``coincidence problem'':
why does the dark energy component happens to become dominant at the
present epoch? Had
this occurred at an earlier epoch, growth of 
structure due to gravitational
instability would be inhibited, and any life form would be
impossible to exist. By introducing a coupling between  
curvature and the quintessence field, it was hoped that the scalar 
field dominance could be
triggered automatically shortly after the Universe becomes matter dominant.
Unfortunately, for the NMC models discussed here, it was found that
the trigger mechanism does not work\cite{NMC,Liddle-Scherrer},
nonetheless, the coupling to gravity could have other interesting 
consequences. Here, I show that NMC models provides a possible
solution of the ``helium problem'' in the big bang nucleosynthesis.

The standard model of Big Bang Nucleosynthesis (BBN) is an 
enormously successful theory. The predicted abundances of the light elements, 
which ranges ten orders of magnitude, were
found to be consistent with observations\cite{OSW00}. In particular, the BBN
prediction of $~^{4}\mathrm{He}$ 
abundance ($Y_p \approx 0.25$) provides the first
evidence of a hot big bang beginning of the Universe. 

\begin{figure}[t]
\epsfxsize=4in 
\epsfbox{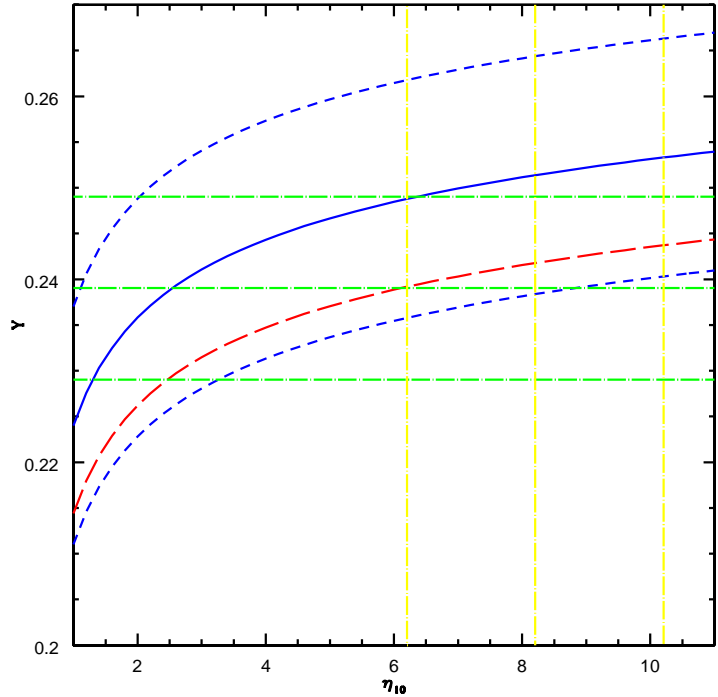} 
\caption{Helium abundance vs. $\eta$. The solid curve is the
prediction of standard BBN with 3 neutrinos, the two short-dashed
curves are standard BBN with 2 and 4 neutrinos, and the long-dashed
curve is the result for a NMC model. The horizontal lines mark the 
center and 2$\sigma$ limits of helium in our ``midway'' approach.
The vertical lines mark the value of $\eta$ determined from 
combined COBE+Boomerang+Maxima data.\label{Y-eta}}
\end{figure}

The $~^{4}\mathrm{He}$ abundance were mostly influenced by two factors,
the expansion rate of the Universe during BBN, 
and the baryon to photon ratio $\eta$  (see Fig.~\ref{Y-eta}). 
The helium abundance increases with the expansion rate for two
reasons. First, BBN starts when
the weak interaction which converts proton to neutron
ceases to be effective. This occurs when
 $\Gamma \sim H$, where $\Gamma, H$ are the
reaction rate and expansion rate, respectively. For a higher $H$ at
a given scale factor $a=1/(1+z)$, BBN starts earlier, when the neutron
fraction is higher. Second, this also meant a shorter interval for the 
neutrons to decay before it is combined in subsequent nuclear fusion.
Both of these two effects enhance the neutron fraction. Since most of 
these neutrons ended up in $~^{4}\mathrm{He}$, 
a faster expanding universe would yield
more helium. Thus, 
once $\eta$ is determined from either deuterium abundance or other
methods such as cosmic microwave background (CMB) anisotropy, 
the $~^{4}\mathrm{He}$ abundance could be used to constrain the expansion rate
during BBN. 

The helium abundance in extragalactic HII (ionized hydrogen)
regions could be obtained by observation of the HeII $\to$ HeI
recombination lines. 
Since $~^{4}\mathrm{He}$ is also produced in stars along with
heavy elements such as Oxygen, it is expected that the primordial
$~^{4}\mathrm{He}$ abundance could be obtained by extrapolation to
zero Oxygen abundance. Using this technique,
Oliver and Steigman obtained \cite{Oliver-Steigman} 
\begin{equation}
Y_p = 0.234 \pm 0.003 (stat.),
\end{equation}
while Izotov and Thuan obtained a higher value\cite{Izotov-Thuan}
\begin{equation}
Y_p = 0.244 \pm 0.002 (stat.).
\end{equation}
Clearly these two data sets are statistically inconsistent with each,
due to large systematic errors. Below, we adopt a midway value of
\begin{equation}
Y_p = 0.239 \pm 0.005,
\label{midway}
\end{equation}
or, $0.229< Y_p < 0.249$ at 95\% C.L.

In order to compare theory with observation, we also need to determine $\eta$.
$\eta$ could be determined from BBN. Burles and
Tytler\cite{Burles-Tytler} obtained  
D/H=$(3.3 \pm 0.25) \times 10^{-5}$, corresponding to a 
lower bound on $\eta$ at $2\sigma$ level
\begin{equation}
\eta_{10} \equiv 10^{10} \eta <6.3.
\end{equation}
CMB anisotropy provides another way of measuring baryon density.
Recently, an analysis of the combined data from Boomerang and Maxima 
yields a higher
baryon density. The best fit model with a flat universe yields\cite{Jaffe}
\begin{equation}
\Omega_b h^2 =0.030 \pm 0.004, \qquad \eta_{10} = 8.2\pm 1.0,
\end{equation}

If we assume that there are three neutrino
species, and adopt $\eta \approx 4.5$ as inferred from the deuterium
abundance, then the standard BBN 
$~^{4}\mathrm{He}$ abundance is in disagreement with the results
of Oliver and Steigman. It is in marginal agreement with 
the ``midway'' result of Eq.~\ref{midway}, but still at high the end. 
If we adopt the $\eta$ inferred from CMB, then even the ``midway''
limit is exceed (see Fig.~\ref{Y-eta}).

Furthermore, in addition to the three standard model neutrinos, 
a sterile neutrino may be needed to explain the results from
neutrino oscillation experiments\cite{neutrino}. If either or both of 
these were confirmed, or if there is any other light particle in the 
Universe, the breach between theory and observation on
$~^{4}\mathrm{He}$ would become even wider.

How could we make a model which produce less helium? If the expansion
of the Universe is slower at the time of BBN, then the helium
abundance is reduced. In the standard BBN model, the expansion rate is 
given by
\begin{equation}
H^2 = \frac{8\pi G}{3} \rho,
\end{equation} 
Thus, $\rho$ becomes greater with the 
introduction of each new particle species. 

In quintessence models, $\rho_{tot} = \rho_f + \frac{1}{2}\dot{\phi}^2 +
V(\phi)$, is it possible to introduce a negative $V$ to reduce $\rho$?
Unfortunately, this would not work. To see this, note that
if a negative potential is introduced, the minima of the potential 
must have $V<0$, the Universe would fall to this potential well. Since
$\rho_f$ is a decreasing function of $a$, sooner or later we would
reach to the point that $\rho_{tot} = 0$, further expansion is not
possible, and the Universe would begin to contract. Such a contraction
of the Universe in the future is not ruled out, however,
if we hope to use a negative potential to reduce helium production, 
the negative potential must become sub-dominant at BBN era, and then 
become dominant well before the current era, which is incompatible
with observation.

With the NMC model introduced in Eq.~\ref{NMCeq}, however, it is possible
to obtain a lower helium abundance, because now we have
\begin{equation}
H^2 = \frac{8\pi G}{3F} \left(\rho+\frac{1}{2}\dot{\phi}^2 + 
V(\phi) -3 \dot{F} H \right).
\end{equation} 
For $F>1$ the value of $H$ could be lowered.

\begin{figure}[t]
\epsfxsize=4in 
\epsfbox{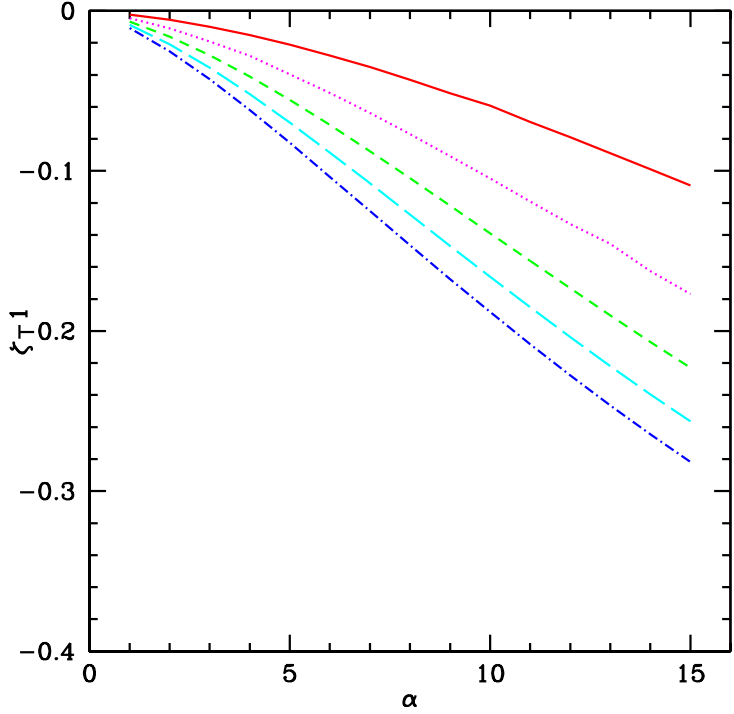} 
\caption{$\zeta-1 $ as a function of $\alpha$, the five curves from
top to bottom are for models with $\xi=0.004, 0.008, 0.012, 0.016,
0.02$.\label{zeta}}
\end{figure}

The Helium abundance in such a model could be estimated. To a first 
approximation, 
\begin{eqnarray}
Y=&&(0.2378+0.0073\ln\eta_{10})(1-0.058/\eta_{10})\\
&&+0.013(N_{\nu}-3)+2\times 10^{-4} (\tau_n - 887).
\end{eqnarray}
The speed up factor 
\begin{equation}
\zeta \equiv H(a)/H_{\mathrm{sm}}(a)
\end{equation}
is related to the neutrino number by
\begin{equation}
\zeta^{2}=1+\frac{7}{43}(N_{\nu}-3).
\end{equation}
So we have 
\begin{equation}
\Delta Y = 0.08(\zeta^2 -1).
\end{equation}

The differential speed up factor $\zeta-1$ for a number of models 
is plotted in Fig.~\ref{zeta}.
As an example, let us consider $\alpha=10$, 
$\xi=0.004$, and $Q_0 =5.5$ which satisfies the solar system limit
$|\xi| <0.022 Q_{0}^{-1}$. The helium abundance could be reduced by
as much as $0.096\%$. In Fig.~\ref{Y-eta}, the helium abundance 
for this NMC model is plotted. It lies much more comfortably within
the allowed range. Alternatively, the helium bound on neutrino number
could be relaxed. If we apply this to the current cosmological limit on
neutrino number \cite{PDG}, which is $1.7<N_{\nu}<4.3$ at $95\%$ C.L., 
the upper limit of $N_{\nu}$ could be lifted to 5. 

In summary, I have shown that in some NMC models, the BBN
helium abundance could be reduced, thus alleviate the marginal
disagreement between theory and observation, and make more room for 
new neutrinos or other new particles.
Whether such a reduction is necessary 
depends on the result of future observations. 

\section*{Acknowledgments}
This work is supported by the US Department of Energy grant
DE-FG02-91ER40690.

\end{document}